\documentclass[prc,superscriptaddress,showpacs,showkeys]{revtex4}
\usepackage{graphicx, latexsym, amsmath, color, multirow, booktabs, ifpdf}
\usepackage{mathrsfs}
\usepackage{amssymb}
\usepackage[section]{placeins}%
\usepackage{amsfonts}

\newcommand{\ff}[1]{\frac{1}{#1}}

\newcommand{\lrl}[1]{\left|#1\right|}

\newcommand{\lrb}[1]{\left(#1\right)}
\newcommand{\lrs}[1]{\left[#1\right]}

\newcommand{\Lrb}[1]{\left\{#1\right\}}

\newcommand{\svec}[1]{{\mbox{\boldmath${ #1}$}}}
\newcommand{\ivec}{\vec}

\newcommand{\figref}[1]{Fig. \ref{#1}}

\begin{document}

\title{The nucleus-nucleus interaction between boosted nuclei}
\author{Wen Hui Long}\email{longwh@lzu.edu.cn}
\affiliation{School of Nuclear Science and Technology, Lanzhou University, 730000 Lanzhou, China}
\affiliation{Department of Physics, Texas A\&M University, Commerce, Texas 75429, USA}
\author{Carlos A. Bertulani}\email{carlos_bertulani@tamu-commerce.edu}
\affiliation{Department of Physics, Texas A\&M University, Commerce, Texas 75429, USA}

\begin{abstract}
The nucleus-nucleus interaction potential has been studied within the relativistic mean field theory. The systematics of the relativistic effects have been investigated by analyzing the relation between the potential and the bombarding energy as a function of the impact parameter. It is shown that the potential barriers are noticeably sensitive to the bombarding energy for a given impact parameter. At large bombarding energies the slope at the potential edge decreases with the impact parameter. Comparisons with a non-relativistic treatment shows that relativistic effects cannot be ignored at bombarding energies around and larger than 100 MeV/nucleon.

\end{abstract}
\pacs{24.10.-i, 24.10.Cn,25.70.Bc}
\keywords{Relativistic many-body, Nucleus-nucleus potential, Elastic scattering}
\maketitle

\section{Introduction}
A consistent theoretical treatment of relativistic many-body systems is still a challenge in many areas of physics. The effects of the Dirac sea imply corrections due to vacuum fluctuations and other features of quantum field theory may also become manifest. An important aspect of special relativity, namely causality, is only verified if sea effects are properly incorporated. Unfortunately, up to now it is proved to be very difficult to find unambiguous experimental evidence for such effects. That might not be the case for static properties of nuclear matter for which a successful project carried out during many years has shown that nuclear energies for the ground and lower excited states of nuclei are well described by a relativistic mean-field treatment of the collection of nucleons \cite{Walecka:1974, Serot:1986, Reinhard:1989, Ring:1996, Bender:2003, Meng:2006}.

Almost nothing has been done to understand the rather relevant problem of the influence of relativity on many-body scattering. Usually, this is accounted quite straightforwardly in quantum field theory for two-body scattering including the sea effects with help of perturbation theory. But the scattering of relativistic composite objects where the compositeness becomes an active part of the scattering process is still a very difficult theoretical problem. This is an obviously important problem in nuclear physics, due to an inherent theoretical difficulty in defining a nuclear potential between many-body relativistic systems. Nonetheless, nuclear potentials are often used in the analysis of nucleus-nucleus scattering at energies of 100 MeV/nucleon and higher. In fact, many rare isotope facilities use reactions at these energies in which relativistic effects are expected to be of the order of 10\% or more for some processes of interest \cite{CB05,OB09,OB10}. For proton-nucleus scattering a successful approach, known as \textquotedblleft Dirac phenomenology\textquotedblright, has been use for many years \cite{AC79}. But for nucleus-nucleus collisions a reasonable account of these features has not yet been accomplished. Due to retardation, an attempt to use a microscopic description for nucleus-nucleus potentials starting from binary collisions of the constituents is not possible as a nucleus-nucleus potential requires a simultaneous interaction between all constituents. In the case of nucleus-nucleus collisions a similar approach as in the Dirac-phenomenology case can be followed up by using a mean field theory.

The nucleus-nucleus potential used in the analysis of elastic scattering experiments is a very complex object which contains information about all possible inelastic processes. The presence of inelastic channels lead to, among other things, the introduction of an imaginary part for the nucleus-nucleus potential.  This potential is usually termed ``optical potential". Although several microscopic models have been developed to calculate the optical potential, due to its complexity, it is often parametrized by a set of phenomenological functions depending on the distance between the center of mass of the interacting nuclei. The parameters set (often with ambiguities) is obtained by fitting elastic scattering data whenever available. Needless to say that for many systems (e.g. most reactions with radioactive beams) the elastic scattering data are unavailable and one has to resort to theoretical constructions of the optical potential. A popular method is to build a folding potential in which the real part of the optical potential is obtained from an integral over the nucleon-nucleon densities weighted by their individual (sometimes density dependent) interactions. The nuclear densities are taken as frozen, with all nucleons interacting simultaneously. A method to deduce the magnitude of the effects of relativity in nucleus-nucleus interactions presented in Refs. \cite{CB05,OB09,OB10} where the nucleus-nucleus potential for a relativistic projectile was modified in a similar way as the scalar part of the electromagnetic interaction. This procedure has shown that relativistic corrections lead to appreciable changes of inelastic processes involving 100 MeV/nucleon projectiles.

In this work, we follow the steps of relativistic mean field theories to construct a relativistic nucleus-nucleus potential. Our goal is to study how relativistic effects would be manifested in the coordinate dependence of a nucleus-nucleus potential. We use a relativistic mean field theory for each nucleus separately for consistency.  The effect we are looking for is obtained by boosting the projectile motion to the target frame of reference (or to the laboratory frame of reference, if both projectile and target are boosted) to analyze how the interaction energy is modified. The interaction energy is defined below as the total energy of the system when the nuclei are at a given finite separation subtracting from it the sum of their individual energies when they are infinitely separated. This is an oversimplification of the problem and it also only allow us to obtain the real part of the interaction potential. In order to study the influence of relativistic effects on elastic cross sections we use, for simplicity, an imaginary part of the optical potential having the same radial dependence as the real part.

\section{Energy density functional for single and two-nucleus systems}

Based on the meson exchange theory, finite nuclear systems can be well described by an energy density functional within the relativistic mean field approach \cite {Walecka:1974, Serot:1986}. In the mean field approach, the nucleons are treated as point-like particles interacting by the exchange of mesons and photons. The energy functional \cite{Serot:1986, Reinhard:1989, Ring:1996, Bender:2003, Meng:2006} for a single nucleus, associated with $\sigma$-, $\omega$-, $\rho$-mesons and photon ($A$) exchange,  is given by ($\hbar=c=1$)
 \begin{equation}\label{Energy_functional}
   E = \int d\svec r \sum_a\bar\psi_a\lrb{-i\svec\gamma\cdot\svec\nabla + M}\psi_a + \ff2\sum_{\phi=\sigma,\omega,\rho, A}\int d\svec r d\svec r' \sum_{ab}\bar\psi_a(\svec r)\bar\psi_b(\svec r')\Gamma_\phi(\svec r, \svec r') D_\phi(\svec r-\svec r')\psi_b(\svec r') \psi_a(\svec r),
 \end{equation}
where the two-body interacting matrices read as
 \begin{subequations}\begin{align}
\Gamma_\sigma(\svec r, \svec r') = &-g_\sigma(\svec r) g_\sigma(\svec r'), \\
\Gamma_\omega(\svec r, \svec r') = &+\lrb{g_\omega\gamma^\mu}_{\svec r} \lrb{g_\omega\gamma_\mu}_{\svec r'},\\
\Gamma_\rho (\svec r, \svec r') = &+\lrb{g_\rho\gamma^\mu\ivec\tau}_{\svec r}\cdot \lrb{g_\rho\gamma_\mu\ivec\tau}_{\svec r'},\\
\Gamma_A (\svec r, \svec r') = &+\frac{e^2}{4}\lrs{\gamma^\mu\lrb{1-\tau_z}}_{\svec r} \lrb{\gamma_\mu\lrb{1-\tau_z}}_{\svec r'}.
 \end{align}\end{subequations}
In the energy functional (\ref{Energy_functional}), $\psi_a$ denotes stationary single particle states, $M$ is the rest mass of nucleon, and $g_\sigma$, $g_\omega$, and $g_\rho$ are the coupling constants with baryonic density dependence \cite{Brockmann:1992, Lenske:1995, Typel:1999, Niksic:2002, Long04}. The propagators of mesons ($\phi=\sigma$, $\omega$, and $\rho$ and $m_\phi$ for the meson mass) and photon ($A$) are given by
\begin{align}
  D_\phi = & \ff{4\pi} \frac{e^{m_\phi\lrl{\svec r-\svec r'}}}{\lrl{\svec r-\svec r'}}, & D_A = & \frac{1}{\lrl{\svec r-\svec r'}}.
\end{align}
In this paper, we use bold types to denote the space vectors and arrows for isospin vectors. Based on the energy functional (\ref{Energy_functional}), the single particle configurations ($\psi_a$) can be obtained from a self-consistent iterative procedure, aiming at describing the physical properties for the ground state of the nucleus such as binding energy, radii, density distributions, etc. \cite{Serot:1986, Reinhard:1989, Ring:1996, Bender:2003, Meng:2006}.

\begin{figure}[htbp]
\includegraphics[width = 0.6\textwidth]{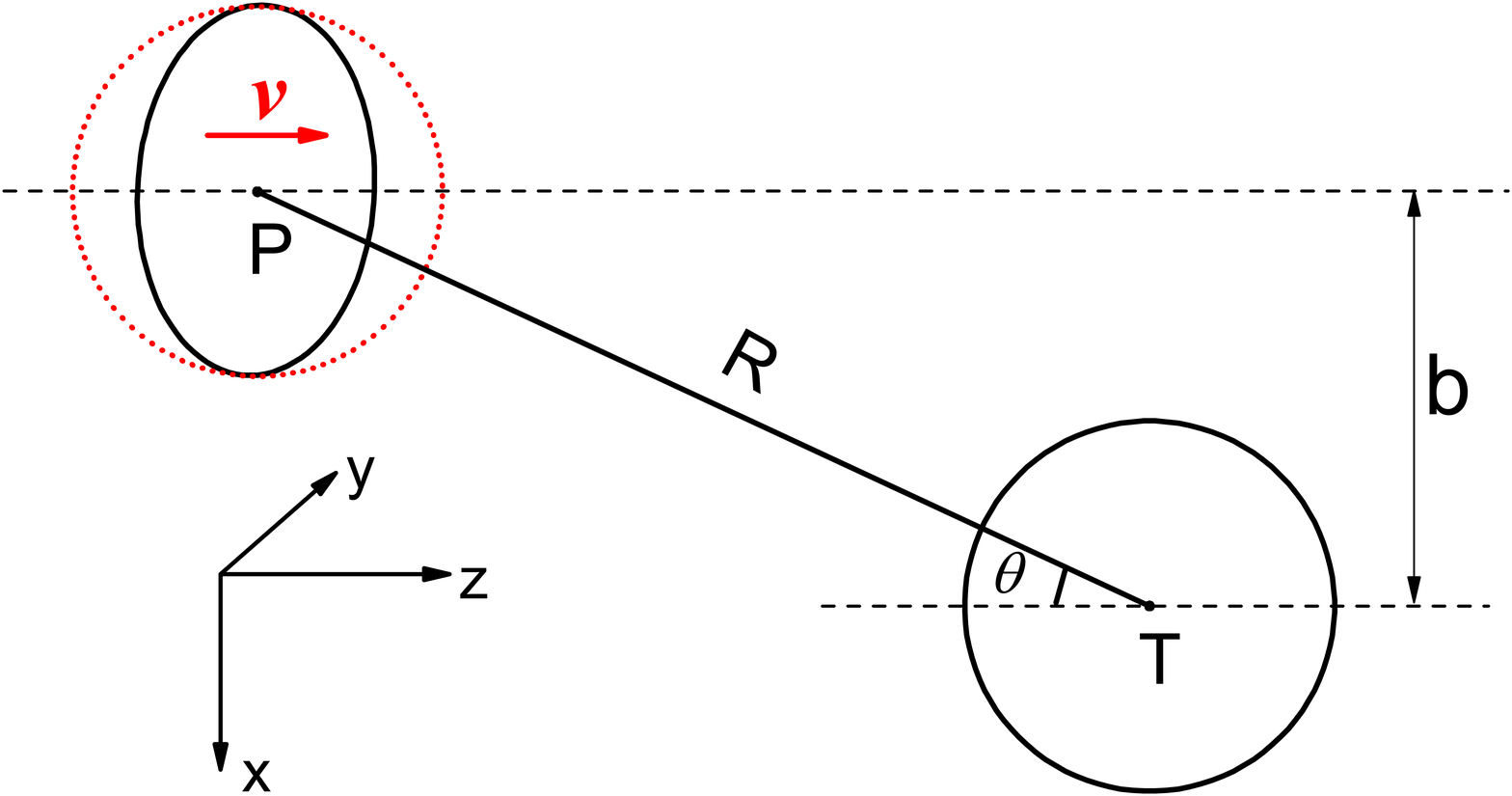}
\caption{(Color online) Two-nucleus system containing rest target and boosted projectile with the velocity $\svec v$. The ellipse represents the boosted projectile density as seen from the target reference frame. $R$ and $b$ denote the distance and impact parameter between two nuclei in the target reference frame.}\label{fig:frame1}
\end{figure}

The relativistic energy functional (\ref{Energy_functional}) described above for a single nucleus can be also extended to describe the interaction between two nuclei, i.e., the interaction between a target and a boosted  projectile. Here we only consider same target and projectile nuclei. As shown in \figref{fig:frame1}, the two-nucleus system consists of a target and a projectile with velocity $\svec v$, separated by a distance $R$.  For simplicity, we assume straight-line dynamics, appropriated for high-energy collisions. A given point in space is denoted by $\svec r_t = (x_t, y_t, z_t)$ in the frame of reference of the target with respect to its center of mass. For a collision with impact parameter $b$ this point can be related to its coordinate $\svec r_p = (x_p, y_p, z_p)$ in the frame of reference of the projectile by means of a Lienard-Wiechert transformation. If  $\svec v$ is taken as the $z$ axis, this transformation reads
 \begin{align}
\label{Frame_C}
x_p =~& x_t + b,& x_t =~& x_p - b,\nonumber \\
y_p =~& y_t,& y_t = ~&y_p,\nonumber \\
z_p =~& \gamma (z_t+R\cos\theta),&z_t = ~& \gamma(z_p - R\cos\theta),
 \end{align}
where the impact parameter is related to the relative distance between the center of mass of the nuclei by $b=R\sin\theta$. The Lorentz factor is $\gamma=(1-\beta^2)^{-1/2}$ and $\beta=v/c$.

The total energy functional for two-nucleus system shown in \figref{fig:frame1} is given by
 \begin{equation}\label{dual}
E(A_t, A_p, \svec v) = E(A_t) + E(A_p, \svec v) + \mathbb E(A_t, A_p, \svec v),
 \end{equation}
where $E(A_t)$ and $E(A_p, \svec v)$ denote respectively the target (with mass number $A_t$) and the projectile (with mass number $A_p$) energy functional, and $\mathbb E(A_t, A_p, \svec v)$ represents the interacting potential energy. For the isolated target and projectile, the energy functional $E(A_t)$ and $E(A_p, \svec v)=E(A_p)$ can be determined from Eq. (\ref{Energy_functional}). Similar as the second term in right hand of Eq. (\ref{Energy_functional}), the interacting one $\mathbb E(A_t, A_p, \svec v)$ can be obtained as
 \begin{equation}\label{two-nucleus}
   \mathbb E(A_t, A_p, \svec v) = \sum_{\phi=\sigma,\omega,\rho, A}\int d\svec r\int d\svec r' \sum_{ab}\bar\psi_{t,a}(\svec r)\bar\psi_{p,b}(\svec r')\Gamma_\phi(\svec r, \svec r') D_\phi(\svec r-\svec r')\psi_{p,b}(\svec r') \psi_{t,a}(\svec r),
 \end{equation}
where the integral over $\svec r$ and $\svec r'$ are done within the same reference frame, e.g., the target inertial frame. The single particle configurations $\psi_{t,a}$ and $\psi_{p,b}$ (subindex $t$ for target and $p$ for projectile) can be determined from the individual self-consistent calculations in the target and projectile frames, respectively. Because the collisions are fast, we adopt a frozen configuration scheme in which the rearrangement effects from the nucleus-nucleus interaction are neglected in determining the single configurations of each nucleus.

Taking $\sigma$ and $\omega$ as example, the corresponding contributions in Eq. (\ref{two-nucleus}) can be written as,
 \begin{align}
\mathbb E_\sigma  = & - \ff\gamma\int d\svec r_t \int d\svec r_p' g_\sigma(\svec r_t)\rho_{s, t}(\svec r_t) D_\sigma(\svec r_t - \svec r_t') \rho_{s, p}(\svec r_p')g_\sigma(\svec r_p'),\\
\mathbb E_\omega  = & +\int d\svec r_t \int d\svec r_p' g_\omega(\svec r_t) \rho_{b, t}(\svec r_t) D_\omega(\svec r_t - \svec r_t') \rho_{b, p}(\svec r_p')g_\omega(\svec r_p').
 \end{align}
The scalar and baryonic densities $\rho_s$ and $\rho_b$ can be obtained as
 \begin{align}
\rho_{s}(\svec r) = & \sum_a\bar\psi_{a}(\svec r)\psi_a(\svec r), & \rho_{b}(\svec r) = \sum_a\bar\psi_{a}(\svec r)\gamma^0\psi_{a}(\svec r),
 \end{align}
where the sums are restricted within one nucleus.

In the interacting energy functional (\ref{two-nucleus}), there exist two types of couplings, the scalar ($\sigma$) and the vectors ($\omega$, $\rho$, and $A$). For a straight-line motion, the Lorentz boots are: $\Lrb{\rho_{b, p}(\svec r_p), 0}\to\Lrb{\gamma\rho_{b, p}(\svec r_p), \gamma\rho_{b, p}(\svec r_p)\svec\beta}$.  The scalar coupling remains invariant under the Lorentz transformations because $d\svec r_t' = d\svec r_p'/\gamma$ and the Lorentz contraction of the longitudinal distance cancels out the Lorentz enhancement of the field. This contrasts with the vector coupling channels where the Lorentz transformation leads to a strong dependence on the projectile velocity.

In energy functional (\ref{two-nucleus}), the coupling constants are functions of density $\rho_v = \sqrt{j^\mu j_\mu}$ and $j^\mu = \bar\psi_t\gamma\psi_t^\mu + \bar\psi_p\gamma^\mu\psi_p$, and the Lorentz transformation from the target to projectile, and the reverse, leads to
 \begin{align}
\rho_v(\svec r_t) = &\sqrt{\rho_{b, t}^2(\svec r_t) + 2\gamma \rho_{b,t}(\svec r_t)\rho_{b,p}(\svec r_p) + \rho_{b,p}^2(\svec r_p)}, \\
\rho_v(\svec r_p) = &\sqrt{\rho_{b, t}^2(\svec r_t) + 2\gamma \rho_{b,t}(\svec r_t)\rho_{b,p}(\svec r_p) + \rho_{b,p}^2(\svec r_p)}.
 \end{align}

The method described above allows us to calculate the interaction potential of the two nuclei in terms of the single-particle states calculated within a relativistic mean field theory. In practice, the calculations are very complicated because the advantage of treating wavefunctions and densities in terms of angular momentum expansions in spherical basis is lost. Due to Eq. \eqref{Frame_C}, the boosted quantities also include Lorentz $\gamma$ factors within the polar angles of the single-particle states for the projectile rendering a very complicated description of their radial and angular dependence.

\section{Results and discussions}

In this work, we take $^{12}$C-$^{12}$C as a typical example of a two-nucleus system in \figref{fig:frame1} and study the nucleus-nucleus interacting potential energy according to the method described above. In the first step, each nucleus is treated in their reference frame independently to obtain the single particle configurations, from which the nucleus-nucleus interacting potential energy is determined via Eq. (\ref{two-nucleus}). The effective interaction PKDD \cite{Long04} is utilized to determine the single particle configurations as well as the nucleus-nucleus potential.

\begin{figure}[htbp]
  \includegraphics[width=0.6\textwidth]{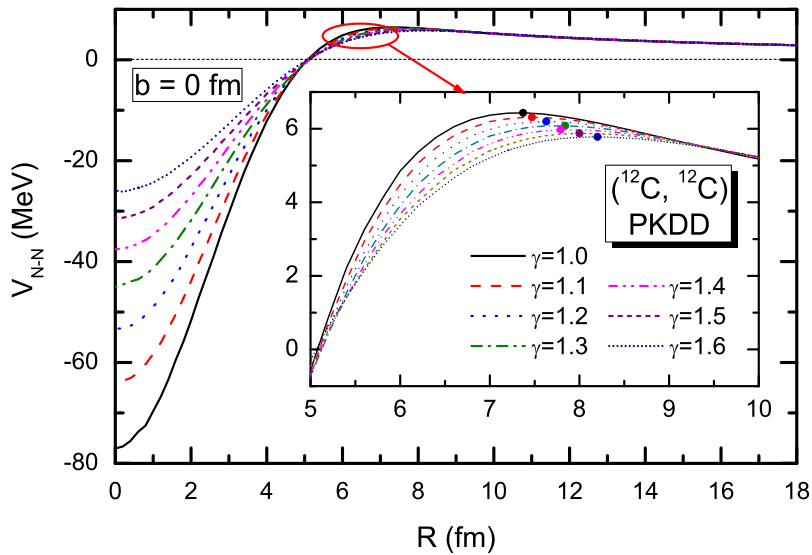}
\caption{(Color online) The interaction potentials (MeV) as functions of the distance $R$ (fm) between rest target and boosted projectile nuclei for impact parameter $b=0$ fm and different values of $\gamma$. The inset presents the potentials from $R=$5 fm to 10 fm with an enlarged scale.}\label{fig:E_RG_b0}
\end{figure}

In \figref{fig:E_RG_b0} we show the nucleus-nucleus potential energies $V(R)=\mathbb E$ [see Eq. (\ref{two-nucleus})] as a function of the distance $R$ for different values of Lorentz factor $\gamma$. The impact parameter is set as $b=0$ fm, and only the surface part of the potential has any useful application. The inset gives the results from $R=5$ to 10 fm with enlarged scale. The nucleus-nucleus potential gets contributions from different fields at different times. First the nuclei approach from a fairly large distance, there exist very little overlap between the target and projectile densities and the potential is mainly contributed by the long range interaction -- the photon field ($A$), which yields the Coulomb barrier. As the distance decreases to, e.g., $R\sim7.5$ fm for $\gamma=1.0$,  the densities  start to overlap and the nucleus-nucleus potential bend toward a negative value. This is due the contribution of the meson fields leading to attraction. When $R \leq 5$ fm, the nucleus-nucleus potential have only net attractions. For the chosen nucleus $^{12}$C, the self-consistent calculation with PKDD gives the neutron, proton and total radii, respectively, as $r_n = 2.26$ fm, $r_p=2.28$ fm, and $r=2.27$ fm. These values (multiplied by 2) are consistent with the evolution of the nucleus-nucleus potential with respect to separation distance $R$, as seen in figure \figref{fig:E_RG_b0}.

The nucleus-nucleus potentials shown in \figref{fig:E_RG_b0} also display a systematical behavior with respect to the Lorentz factor $\gamma$ which represents the boosted energy of the projectile, roughly $(\gamma-1)M$ per nucleon. With increasing bombarding energy, the nucleus-nucleus potential well becomes shallower, as seen in \figref{fig:E_RG_b0}. The point where the potential becomes attractive does not change appreciably with the bombarding energy. But, as shown in the inset, the potential barrier does have a strong dependence on the bombarding energy. This hints to a destructive additional (because of the boost) cancellation between the attractive meson fields and the repulsive photon field. These additional cancellations also tend to lead to a shallower potential.  The filled circles in the inset show the tendency of this effect to stretch the nucleus-nucleus potential outwards as the energy increases.

\begin{figure}[htbp]
  \includegraphics[width=0.90\textwidth]{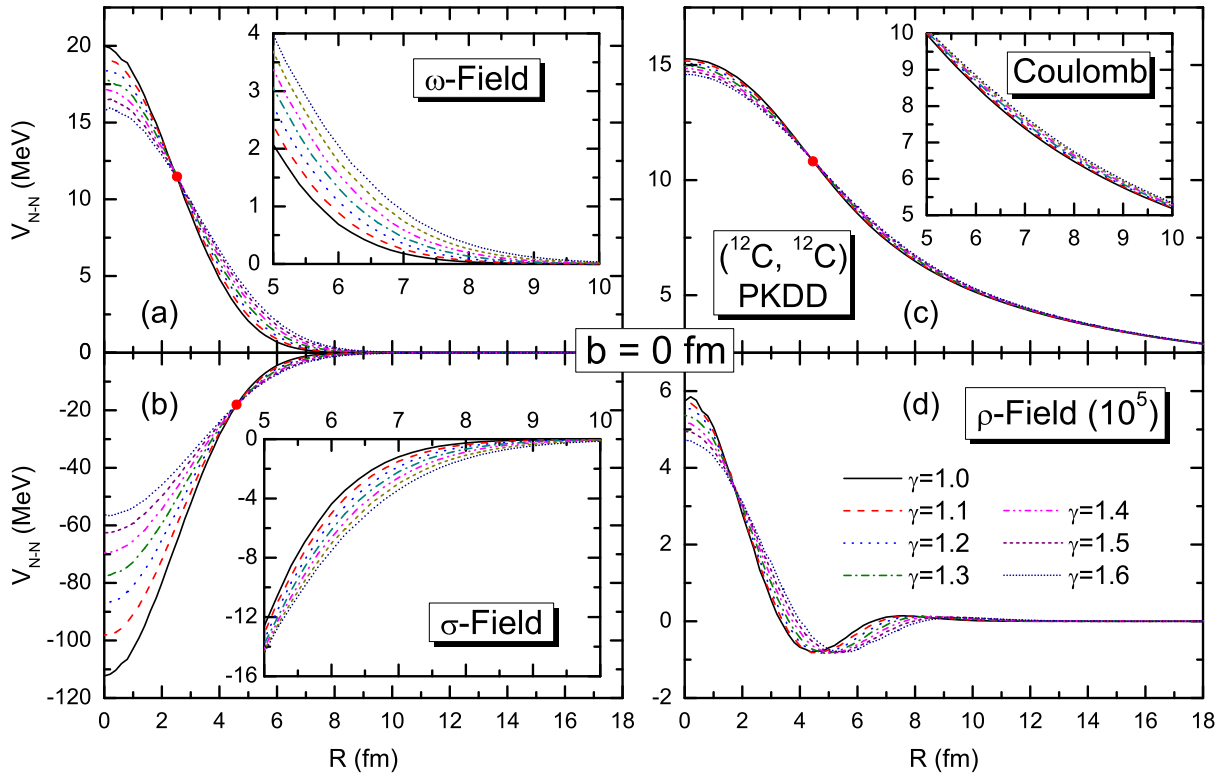}
  \caption{(color online) Contributions to the interaction potentials from the $\omega$ (a), $\sigma$ (b), Coulomb (c), and $\rho$ (d) fields are shown as functions of distance $R$ (fm) with different values of $\gamma$ for the same two-nucleus system shown in \figref{fig:frame1}. The insets show the details from $R=5$ fm to 10 fm with enlarged scales.}\label{fig:E_RG_40}
\end{figure}

In order to have better understanding about the systematics of the nucleus-nucleus potentials, we show in \figref{fig:E_RG_40} the detailed contributions from four coupling channels, (a) $\omega$-vector, (b) $\sigma$-scalar, (c) Coulomb vector, and (d) $\rho$-vector couplings. The corresponding insets present the results from $R=5$ to 10 fm with enlarged scales. In \figref{fig:E_RG_40} (d), it is clearly seen that the contribution from the $\rho$ field is very tiny since in $^{12}$C the nucleons are isospin saturated. Among the contributions to the nucleus-nucleus potential, the attraction mainly originate from the $\sigma$ scalar coupling, whereas the repulsion is mainly due to the $\omega$ and Coulomb fields. Starting from a fairly large separation distance,  the sole contribution photon field yields and increasing Coulomb repulsion. At $R\sim10$ fm the $\sigma$ and $\omega$ fields start to act with distinct contributions to turn the Coulomb repulsion into nuclear attraction. From \figref{fig:E_RG_40}(a) and (b), one sees that the $\omega$ and $\sigma$ contributions increase rapidly when the target and projectile approach. This is mainly due to the short range character of the meson exchange interaction.

As seen from the inset of \figref{fig:E_RG_40} (c), the photon field has a rather weak dependence on the boosted energy whereas a stronger dependence is found from the contributions from $\sigma$ and $\omega$ fields. As shown in the insets of \figref{fig:E_RG_40} (a) and (b), the repulsive ($\omega$ field) and attractive ($\sigma$ field) increase with the boosted energy considerably. When the target and projectile approach a close distance, the curves with different bombarding energies tend to cross each another at some point, denoted by red circles in \figref{fig:E_RG_40}. In fact these are turning points, where the trend of the bombarding energy dependence is just reversed. Similar turning points can be also found in \figref{fig:E_RG_b0}, where the nucleus-nucleus potential becomes attractive.

\begin{figure}[htbp]
  \includegraphics[width = 0.90\textwidth]{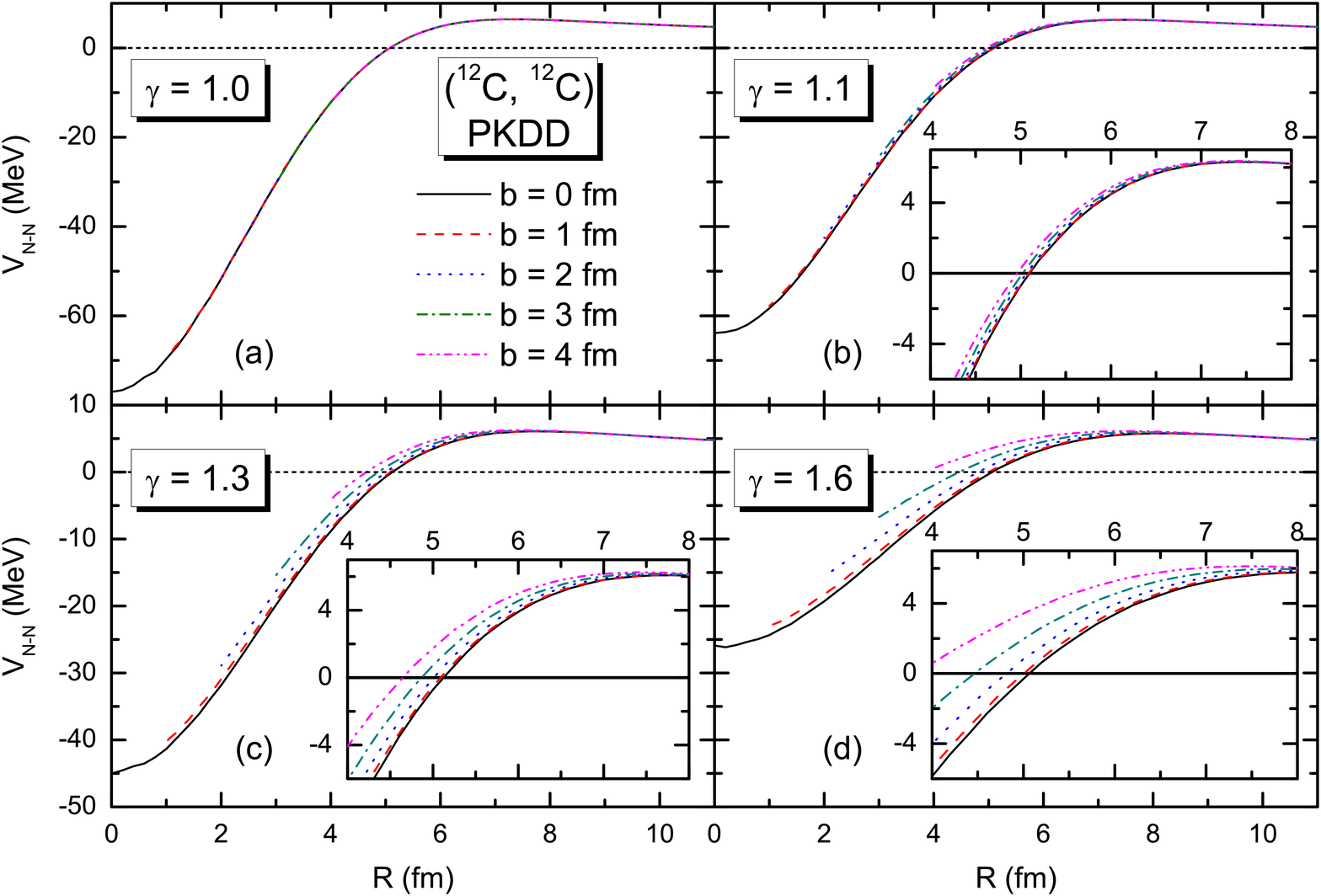}
  \caption{(Color online) The nucleus-nucleus interaction potentials (in  MeV) are shown as function of the distance $R$ between the target and projectile nuclei for different values of $\gamma$  [(a) $\gamma=1.0$; (b) $\gamma=1.1$; (c) $\gamma=1.3$; and (d) $\gamma=1.6$] and different impact paramters $b$ (fm). The insets show the results from $R=3.5$ fm to 8 fm with enlarged scales.}\label{fig:E_RG_BA}
\end{figure}

In \figref{fig:E_RG_b0} and \figref{fig:E_RG_40}, the impact parameter $b$ is set to zero. In \figref{fig:E_RG_BA} we show the nucleus-nucleus potential as a function of the separation distance $R$ with the impact parameter $b=0, 1, 2, 3$, and 4 fm. For comparison, the results with four bombarding energies are shown in \figref{fig:E_RG_BA}, (a) $\gamma=0$, (b) $\gamma=1.1$, (c) $\gamma=1.3$, and (d) $\gamma=1.6$. As seen from \figref{fig:E_RG_BA} (a) and (b), the nucleus-nucleus potentials with lower bombarding energy are not sensitive to the impact parameter. With increasing bombarding energy, the potentials show a strong dependence on the impact parameter. As shown in \figref{fig:E_RG_BA} (c) and (d), the slope of the nucleus-nucleus potential at large distances becomes smaller with increasing impact parameter. It is also demonstrated by the insets in \figref{fig:E_RG_BA}.

In order to asses the relativistic modifications of the nucleus-nucleus potential on scattering observables, we consider the elastic scattering cross section of spherically symmetric nuclei. At high energies the scattering amplitude is well described by the eikonal approximation,
\begin{equation}
f(\vartheta)= f_C(\vartheta) + ik\int_0^\infty J_0(qb) \exp\lrs{i\chi_C(b)} \Lrb{1 - \exp\lrs{i\chi_N(b)}} b db
\label{eikftheta}
\end{equation}
where $J_0$ is the Bessel function of zero-th order, $q=2k\sin(\vartheta/2)$ is the momentum transfer for the scattering angle $\vartheta$ and projectile momentum $\hbar k = \mu v$. The reduced mass is $\mu = \gamma m_0/(1+\gamma)$, where $m_0$ is the rest mass of $^{12}$C, and $v =  c\sqrt{(\gamma - 1)/\gamma}$. The Coulomb-scattering amplitude reads
\begin{equation}
f_C(\vartheta) = \frac{Z_1Z_2e^2}{2\mu v^2\sin^2(\vartheta/2)}\exp\Lrb{-i\eta\ln\lrs{\sin^2(\vartheta/2)} + i\pi + 2i\phi_0},
\end{equation}
where $\eta = 2Z_1Z_2e^2/\hbar v$, and $\phi_0 = {\rm arg} \Gamma\lrb{1+ \eta/2}$.

The eikonal phase $\chi_N$ and $\chi_C$ in Eq. (\ref{eikftheta}) are given by
\begin{align}
\chi_N(b)=&-\frac{1}{\hbar v}\int_{-\infty}^{\infty}\,U_{opt}(b,z') dz', & \chi_C(b) = & \frac{2Z_1Z_2e^2}{\hbar v} \ln(kb), \label{eikphase}
\end{align}
where $\chi_C(b)$ represents the Coulomb eikonal phase.

As mentioned before, the imaginary part of the optical potential has to be included by hand. In terms of the real part of the nucleus-nucleus potential $V(\svec r)$ ($\svec r=({\svec b},z)$ and $r=\sqrt{b^2+z^2}$) calculated above, the nuclear optical potential will be written as
 \begin{equation}
U_{opt}(\svec r)  = V (\svec r) + i\lambda V(\svec r).
 \end{equation}
The coefficient $\lambda$ is set arbitrarily to 0.8 in the calculations. The elastic cross section is then given by $d\sigma/d\Omega=|f(\vartheta)|^2$. As usual, the elastic cross sections will be described in units of the the Rutherford differential cross section
\begin{equation}
\frac{d\sigma}{d\Omega} = \lrl{f_{\text{Ruth.}}(\vartheta)}^2 = \lrb{\frac{Z_1Z_2e^2}{2\mu v^2}}^2 \frac{1}{\sin^4\lrb{\vartheta/2}}.
\end{equation}

\begin{figure}[htbp]
\includegraphics[width=0.5\textwidth]{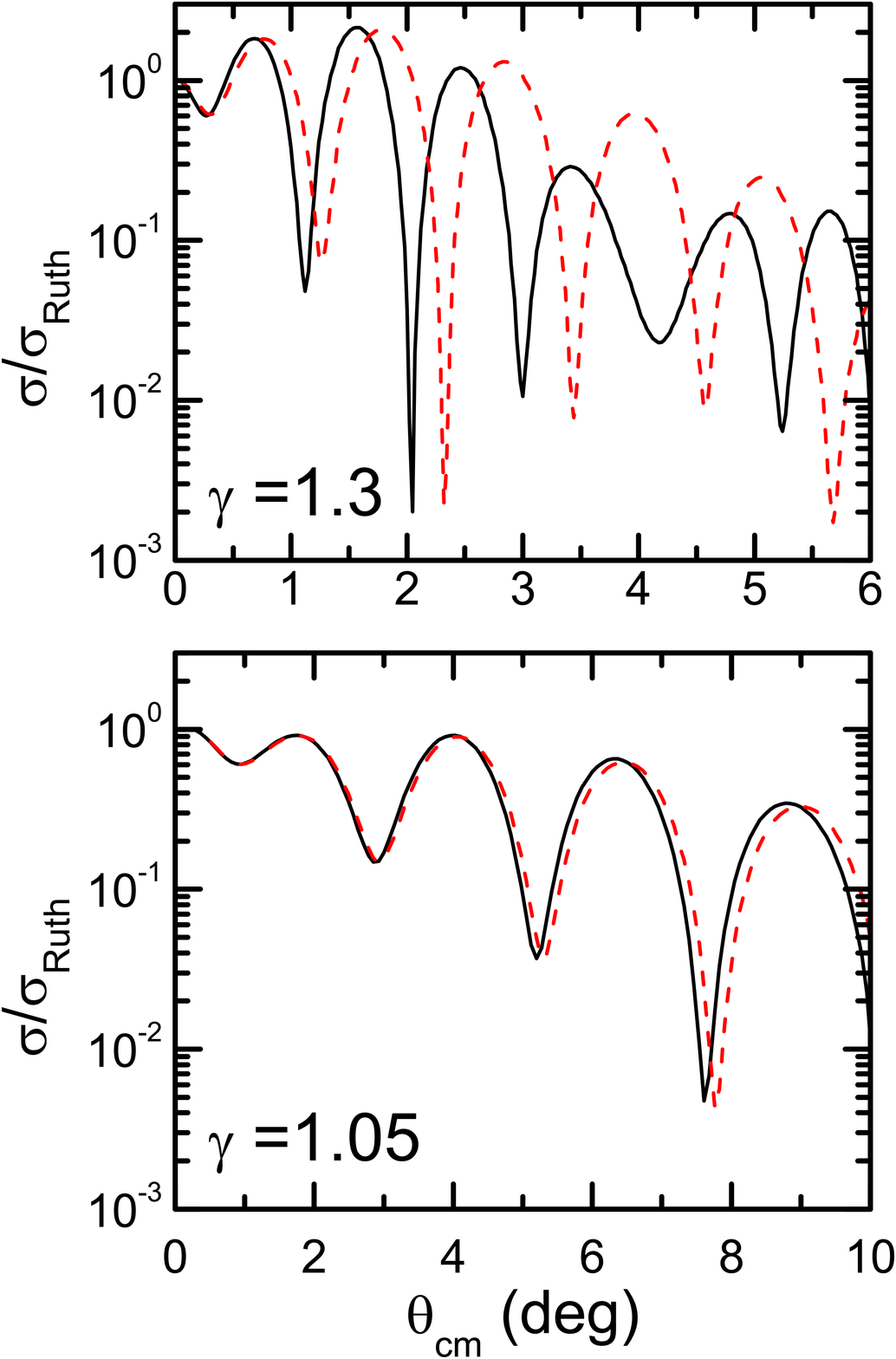}
\caption{(Color online) Elastic cross sections $d\sigma/d\Omega=|f(\vartheta)|^2$ in units of Rutherford cross sections $\lrl{f_{\text{Ruth.}}(\vartheta)}^2$ as functions of the scattering angle $\vartheta$, with different values of Lorentz factor $\gamma$. The dashed lines represent the results with non-relativistic reduction.}\label{fig:R30}
\end{figure}

Our results for $^{12}$C+$^{12}$C at $\gamma=1.05$ (approximately 50 MeV/nucleon) and $\gamma = 1.3$ (approximately 300 MeV/nucleon) are shown in \figref{fig:R30} (solid lines). We compare the cross sections with those calculated with non-relativistic dynamics are shown in \figref{fig:R30} by dashed lines. On observes that the minima of the elastic cross sections are shifted to smaller values of the scattering angle. This is due to the smaller effective radius of the nucleus-nucleus potential when relativistic effects are included, as observed in \figref{fig:E_RG_b0}-\figref{fig:E_RG_BA}. For a discussion of the effects of size and diffuseness of optical potentials on scattering observables, see, e.g., Ref. \cite{BG10}. The  diffuseness of the potentials also increase with the inclusion of relativistic effects. This is shown manifest in Fig.  \figref{fig:R30} by a (slightly) faster decrease of the cross section as a function of angle.

In summary, we have investigated the effects of relativity on nucleus-nucleus potentials within the relativistic mean field theory. The relativistic effects have been studied by analyzing the dependence of the potentials upon bombarding energies and impact parameters. It is found that for a given impact parameter the Coulomb barrier is softened with increasing bombarding energy. For large bombarding energies, the potential edge becomes increasingly flat with increasing impact parameters which indicates that the target and projectile have to get closer to compensate for more attraction. The studies carried out here are exploratory in the sense that a consistent theory of a nucleus-nucleus potential including relativity requires a much more elaborated theory. Such a theory is missing in the literature. Many of nucleus-nucleus experiments at energies of 100 MeV/nucleon and above are presently been carried out around the world with the goal to extract spectroscopic information on rare nuclear species \cite{BG10}. Our work shows that some modifications on the values of the extracted spectroscopic quantities might occur due to the relativistic dynamics missing in most methods used to construct a nucleus-nucleus potential, such as the folding models.

\section{acknowledgments}
The authors acknowledge support by the US Department of Energy under Grant Nos. DE-FG02-08ER41533 and DE-FC02-07ER41457 (UNEDF, SciDAC-2) and the Research Corporation. This work is also partially supported by the National Natural Science Foundation of China under Grant No. 11075066, and the Fundamental Research Funds for Central Universities under contract No. lzujbky-2010-25.


%

\end{document}